\documentclass[twocolumn,prl,aps,nobalancelastpage,superscriptaddress]{revtex4}

\usepackage{graphicx}
\usepackage{bm}
\begin{document}
\title{Fermiology and electronic homogeneity of the superconducting overdoped cuprate Tl$_2$Ba$_2$CuO$_{6+\delta}$
revealed by quantum oscillations}

\author{A.~F.~Bangura}
\affiliation{H. H. Wills Physics Laboratory, University of Bristol, Tyndall Avenue, BS8 1TL, UK.}
\author{P.~M.~C.~Rourke}
\affiliation{H. H. Wills Physics Laboratory, University of Bristol, Tyndall Avenue, BS8 1TL, UK.}
\author{T.~M.~Benseman}
\affiliation{Physics Department, Cavendish Laboratory, University of Cambridge, J. J. Thomson Avenue, Cambridge, CB3 0HE, UK.}
\author{M.~Matusiak}
\affiliation{Physics Department, Cavendish Laboratory, University of Cambridge, J. J. Thomson Avenue, Cambridge, CB3 0HE, UK.}
\author{J. R.~Cooper}
\affiliation{Physics Department, Cavendish Laboratory, University of Cambridge, J. J. Thomson Avenue, Cambridge, CB3 0HE, UK.}
\author{N. E.~Hussey}
\affiliation{H. H. Wills Physics Laboratory, University of Bristol, Tyndall Avenue, BS8 1TL, UK.}
\author{A.~Carrington}
\affiliation{H. H. Wills Physics Laboratory, University of Bristol, Tyndall Avenue, BS8 1TL, UK.}

\date{\today}

\begin{abstract}
We report an angular quantum oscillation study of Tl$_2$Ba$_2$CuO$_{6+\delta}$ for two different doping levels ($T_c = 10$\,K and 26\,K)
and determine the Fermi surface size and topology in considerable detail. Our results show that Fermi liquid behavior is not confined to
the edge of the superconducting dome and is robust up to at least $T^{\rm max}_c$/3.5. Superconductivity is found to survive up to a larger
doping $p_c = 0.31$ than in La$_{2-x}$Sr$_x$CuO$_4$. Our data imply that electronic inhomogeneity does not play a significant role in the
loss of superconductivity and superfluid density in overdoped cuprates, and point towards a purely magnetic or electronic pairing
mechanism.
\end{abstract}

\maketitle The evolution of the electronic structure with carrier concentration is crucial for understanding the origin of high temperature
superconductivity. In the well-studied hole-doped cuprate La$_{2-x}$Sr$_x$CuO$_4$ (LSCO), antiferromagnetism is suppressed beyond $x$ =
0.02 and superconductivity emerges at $x$ = 0.05. The superconducting (SC) transition temperature $T_c(x)$ is approximately an inverted
parabola maximizing at $x \sim$ 0.16 before vanishing at $x$ = 0.27 \cite{Takagi89}. Over much of this phase diagram the physical
properties are substantially different from those of conventional metals and the range of validity of a Fermi liquid picture, where all
quasiparticles are well defined, is unclear.

In fully-oxygenated LSCO, it is usually assumed that  $x$ equals $p$, the number of added holes/CuO$_2$ unit. Tallon {\it et al.} have
argued that $T_c$ follows a universal dependence on $p$ for  {\it all} hole-doped cuprate families; $T_c$/$T_c^{\rm max}$ = 1 - 82.6($p$ -
0.16)$^2$) \cite{Tallon95}. In many families however, the precise doping level is difficult to determine. Moreover, it has been suggested
that hole-doping in cuprates is intrinsically inhomogeneous on a length scale of a few unit cells \cite{Uemura01, Cren00, McElroy05,
EmeryKivelson93}, with $p$ merely being  a global average of the ensemble. NMR experiments on YBa$_2$Cu$_3$O$_{7-\delta}$ however, suggest
that (static) phase separation is {\it not} a generic property of underdoped (UD) or optimally doped (OP) cuprates \cite{Bobroff02}.  This
is supported by analysis of heat capacity data for Bi$_2$Sr$_{2}$CaCu$_{2}$O$_{8+\delta}$ over a wide range of $p$, and of NMR data for
YBa$_2$Cu$_4$O$_{8}$ and Ca-doped YBa$_2$Cu$_3$O$_{7-\delta}$ \cite{Loram04}. For overdoped (OD) cuprates, the applicability of the phase
separation picture is still debated. Experimentally, the ratio of the superfluid density $n_s$ to the carrier effective mass $m^*$ is found
to decrease with increasing $p$ \cite{Niedemayer93, Uemura93}. This so-called \lq boomerang' effect has been attributed either to
pair-breaking in an homogeneous electronic state \cite{Niedemayer93}, or to spontaneous phase separation into hole-rich (non-SC) and
hole-poor (SC) regions \cite{Uemura01, Uemura93,Tanabe05,Wang07}.

Here, we report a detailed study of the de Haas-van Alphen (dHvA) effect in Tl$_2$Ba$_2$CuO$_{6+\delta}$ (Tl2201) single crystals with two
different $T_c$ values. Our results show that a generalized Fermi-liquid picture extends into the high-$T_c$ phase of OD cuprates and is
not confined to the edge of the SC dome. We show that OD Tl2201 has a highly homogenous electronic state on the scale of the mean-free-path
($\ell_0 \gtrsim 400$\,\AA). Our precise determination of the Fermi surface (FS) volume reveals that superconductivity in Tl2201 survives
up to $p_c$ = 0.31 significantly beyond that inferred for LSCO. The lack of nanoscale inhomogeneity implies that the rapid loss of
superfluid density with overdoping is likely due to pair breaking driven by a weakening of the pairing interaction. Finally, the dHvA mass
does not change much with $p$ and seems to correspond to an overall band-narrowing rather than strong renormalization near the Fermi level.
These two facts tend to rule out pairing mechanisms involving low frequency bosons for these overdoped cuprates.

Tl2201 crystals were grown using a self-flux method \cite{Tyler97} and annealed in flowing oxygen at various temperatures to achieve a
range of $T_c$ values \cite{Tyler97}. Torque magnetization was measured using a piezo-resistive microcantilever in a $^3$He cryostat in the
45 Tesla hybrid magnet in Tallahassee. The tetragonal crystal structure and orientation were determined by x-ray diffraction.

\begin{figure}
\includegraphics*[width=0.95\linewidth,clip]{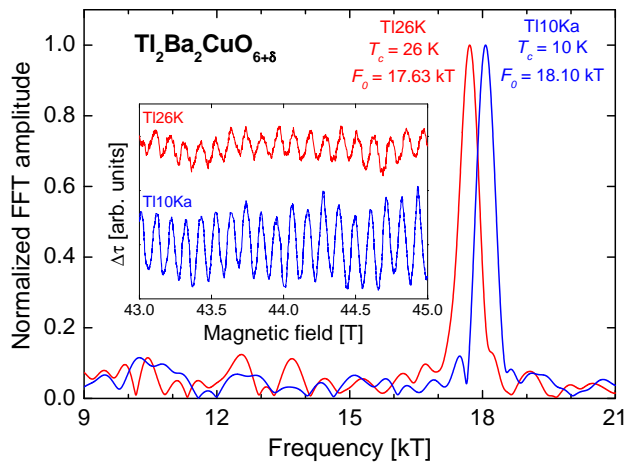}
\caption{(color online) Fast Fourier transform of torque data between 38\,T and 45\,T  for Tl26K (red) and Tl10Ka (blue) samples. The inset
show the raw data for the two samples.} \label{Fig1}
\end{figure}

In the inset of Fig.~\ref{Fig1} we show torque data for samples with two different doping values ($T_c = 10$\,K and 26\,K) after
subtracting a 3$^{\rm rd}$ order polynomial fit to raw data obtained at $T = 0.4$\,K.  (Hereafter, crystals are identified by their $T_c$
values prefixed by the letters \lq Tl'.) Clear dHvA oscillations are seen in both cases. As shown in the main panel, fast Fourier
transforms (FFT) of the raw data show a single sharp peak in the spectra. Note that the slow period oscillations which are evident in the
raw data are not intrinsic. The clear shift of the dHvA frequency $F$ to a lower value for the higher $T_c$ sample confirms that the
oscillations do not arise from any part of the sample which is not superconducting. Although experiments were made on a large number of
samples with different $T_c$ (up to 60\,K) in fields up to 70\,T (in Toulouse), clear oscillations were only seen on three crystals and
weak signals in one other. Samples with low impurity scattering, high crystallographic quality and high doping homogeneity are needed to
observe dHvA oscillations arising from such large Fermi surface sheets. For example, a factor two decrease in $\ell_0$ from the present
values would result in a reduction in the dHvA amplitude of $10^5$ burying the signal firmly into the noise.

Fitting the angle dependence $F(\theta)$ to the expression for a two-dimensional (2D) FS (= $F_0/\cos{\theta}$), we obtain a fundamental
frequency $F_0$ of 18.10(3)\,kT and 18.00(1)\,kT for  Tl10Ka and Tl10Kb respectively, in agreement with a previous study in pulsed fields
\cite{Vignolle08, angleerrornote}, and 17.63(1)\,kT for Tl26K.  Note that the width of the FFT peak in Fig.\ \ref{Fig1} is determined by
the field range and is not a measure of the accuracy in $F(\theta)$. $F_0$ gives directly the extremal cross-sectional area $\mathcal{A}$
of the Fermi surface via the Onsager relation $\mathcal{A} = 2 \pi e F_0/ \hbar$.  As $\mathcal{A}$ does not vary appreciably with $k_z$
(see later) we can determine the carrier concentration $(1+p)$ via Luttinger's theorem, $1+ p = 2\mathcal{A}/(2\pi/a)^2$ (where $a =
3.86$\,\AA\ is the in-plane lattice parameter). From the frequencies, we obtain hole dopings, $p$,  of 0.304(2), 0.297(1) and 0.270(1)
respectively for Tl10Ka, Tl10Kb and Tl26K. These values are consistent with the measured zero-temperature Hall number $n_{\rm H}(0) =
1.28(6)$ /Cu for a $T_c$=15\,K sample \cite{Mackenzie96b}.

\begin{figure}
\includegraphics*[width=0.9\linewidth,clip]{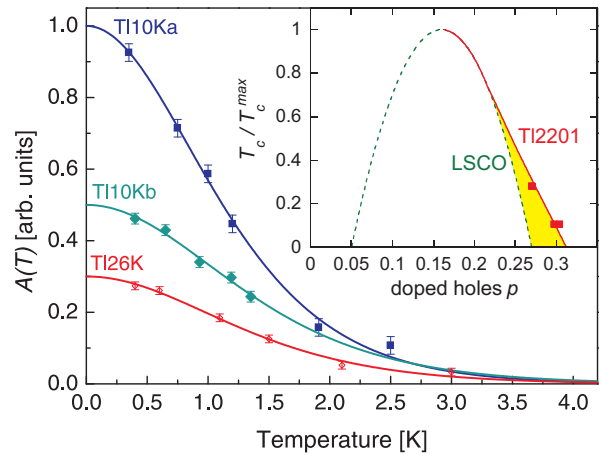}
\caption{(color online) Temperature dependence of the oscillatory torque amplitudes for the three crystals used in this study, fitted to
$R_T = \chi/\sinh(\chi)$, with $\chi = 14.695 m^* T / B \cos \theta$. Field ranges are 40--45\,T for Tl10Ka (squares), 42--45\,T for Tl10Kb
(diamonds) and 43--45\,T for Tl26K (circles). The inset shows the superconducting $T_c$ versus $p$ dome for LSCO (dashed line)
\cite{Tallon95} and Tl2201 (solid line and squares). The (yellow) shaded area represents the proposed region of suppressed
superconductivity in OD LSCO.} \label{Fig2}
\end{figure}

We determine the quasiparticle effective mass by analyzing the $T$-dependence of the dHvA amplitude $A$ using standard Lifshitz-Kosevich
(LK) theory \cite{Shoenberg}.  Results are shown for the three different crystals in Figure \ref{Fig2}.  $A(T)$ is well fitted by the
standard LK formula showing that possible deviations from the usual thermal population factor (the Fermi function) arising from (marginal)
non-Fermi liquid effects \cite{Wasserman} are not apparent for temperatures above $\sim$ 350\,mK.  From the fits, we obtain effective
masses $m^*$ of 5.8(2), 4.9(2) and 5.0(2) $m_e$ at $\theta = 0^{\circ}$ for Tl10Ka, Tl10Kb and Tl26K respectively. To check for consistency
and any field dependence of the dHvA mass (common in heavy Fermion systems, where $m^*$ is strongly enhanced by spin-fluctuations) we
compare these values to the zero field electronic specific heat.  For a 2D metal, the Sommerfeld coefficient $\gamma$ = $(\pi k_{\rm
B}^2N_Aa^2/3\hbar^2)m^*$ (where $k_{\rm B}$ is the Boltzmann constant and $N_A$ Avogadro's constant) \cite{Mackenzie96a}. Accordingly, we
obtain $\gamma$ = 7.1(4)\,mJ mol$^{-1}$K$^{-2}$ for both dopings, in excellent agreement with the almost $p$-independent value of 7(1) mJ
mol$^{-1}$K$^{-2}$ found from direct measurement of polycrystalline Tl2201 \cite{Wade94}.

Comparison with the band mass $m_b \sim 1.7$\,$m_e$, given by density functional theory band calculations \cite{RourkeNJP, Sahrakorpi}
(with the FS area adjusted using the virtual crystal approximation to match the present doping), reveals a significant enhancement due to
interaction effects ($m^*/m_b\approx$ 3) that is constant (within our uncertainty) up to at least $\sim$ $T^{\rm max}_{\rm c}$/3.5.  This
renormalisation is $\sim 20$\% less than the band-width renormalisation (which corresponds to $m^*/m_e \sim 6.6$) found by angle-resolved
photoemission spectroscopy (ARPES) for OD Tl2201 ($T_c$ = 30\,K) \cite{Plate05}. This implies that the renormalisation predominantly arises
from correlation-induced band-narrowing and that any further renormalisation close to the Fermi level $E_F$ from interaction with low
energy boson modes is minimal.

\begin{figure}
\includegraphics[width=0.95\linewidth,clip]{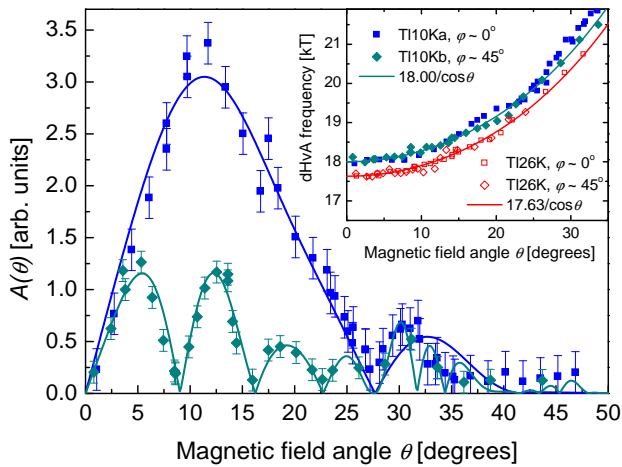}
\caption{(color online) Polar angle dependence of the oscillatory torque amplitude for Tl10Ka at $\varphi \sim 0^{\circ}$ (squares) and
Tl10Kb at $\varphi \sim 45^{\circ}$ (diamonds). The solid line fits to $A(\theta)$ are explained in the text \cite{actualangles}. The
insert shows the polar angle dependence of the dHvA frequencies for all three samples.} \label{Fig3}
\end{figure}

Details of the FS topology and spin-susceptibility can be determined by measuring the dependence of the dHvA frequency and amplitude on the
magnetic field angle with respect to the $c$-axis ($\theta$) and the Cu-O bond direction ($\varphi$). Fig.~\ref{Fig3} shows the FFT
amplitudes extracted from field sweeps at $T = 0.45$\,K for Tl10Ka (at $\varphi \sim 0^{\circ}$ and field range 42--44\,T) and Tl10Kb (at
$\varphi \sim 45^{\circ}$ and field range 42--45\,T). Qualitatively similar behavior was observed for the 26\,K sample \cite{RourkeNJP}.
For $\varphi \sim 0^{\circ}$, a broad peak is observed, followed by a minimum that occurs at $\theta \sim$\,27.5$^{\circ}$. For $\varphi
\sim 45^{\circ}$, the behavior is strikingly different with several clear minima below this angle.

In a quasi-2D metal minima in  $A(\theta,\varphi)$ can occur due to beats between FS sections with nearly equal area. Following Bergemann
{\it et al.} \cite{Bergemann03}, we expand the Fermi surface using cylindrical harmonics compatible with the body centered tetragonal
crystal symmetry
\begin{equation}
  k_{\rm F}(\varphi,\kappa_z) = \displaystyle\sum_{m,n}k_{mn} {\rm cos}(n\kappa_z) \times\left\{\begin{array}{l}{\rm cos}(m\varphi)~(m=0,4,8...)\\{\rm sin}(m\varphi)~(m = 2,6,10...)\end{array}\right.
\label{outplane0}
\end{equation}
As the FS is close to 2D we determined the dHvA frequencies and amplitudes from FFTs of the oscillatory magnetisation calculated using the
full superposition integral, $M\propto \int_0^{2\pi} dk_z \sin(\hbar \mathcal{A}(k_z)/eB)$ (where  $\mathcal{A}(k_z)$ is determined from
numerical integration of  $k_F(\varphi,\kappa_z)$ including spin-splitting) rather than the more usual extremal approximation
\cite{Bergemann03}. As a guide we use the same components ($k_{00}$, $k_{40}$, $k_{21}$, $k_{61}$ and $k_{101}$) that were used to map out
the FS topology of OD Tl2201 by angle-dependent magnetoresistance (ADMR) \cite{Hussey03, actualvalues}. The Fermi surface volume (and hence
$1+p$) is determined by $k_{00}$ alone and this is precisely determined by the mean dHvA frequency. For $\varphi=0$, the $n = 1$ warpings
do not give rise to beats for any $\theta$ and hence at low $\theta$, where spin-splitting effects are negligible, data for the field
dependence of the amplitude for this direction can be simply related to $\ell_0$, giving values of 420(10)\,\AA~ for sample Tl10Ka and
522(100)\,\AA~ for Tl26K. These $\ell_0$ values show that  quasiparticle decoherence hardly changes as the doping is decreased towards
$T_c^{\rm max}$. This conclusion is supported by the relatively minor difference in absolute dHvA amplitude between the 10\,K and 26\,K
samples.

For $\varphi \sim 45^\circ$, the minima observed below $\theta=25^\circ$ are determined by $k_{21}$, the dominant component of the $c$-axis
warping and a parameter that cannot be determined directly by ADMR or ARPES. Our data show that $k_{21} = 0.00170(5)$\,\AA$^{-1}$ and
$0.00125(5)$\,\AA$^{-1}$ for $T_c = 10$\,K and 26\,K respectively, corresponding to a resistivity anisotropy $\rho_c/\rho_{ab}$ of
$3.3(2)\times 10^3$ and $6.2(5)\times 10^3$ (assuming isotropic scattering) and a significant increase in the electrical anisotropy as
$T_c^{\rm max}$ is approached.

Minima in $A(\theta,\varphi)$ can also occur due to spin-splitting, and a `spin-zero' is expected when the spin-up and spin-down FS areas
differ by a half-integral number of Landau quanta \cite{Shoenberg}.  The location of these `spin-zeros' is determined by the spin-mass
$m_{sus}$, which is generally not the same as $m^*$ \cite{Bergemann03}.  As  no minima due to FS warping are expected for $\varphi=0$ we
attribute the minimum that occurs at $\theta \sim$\,27.5$^{\circ}$ to such spin-splitting.  Because only one minimum is observed $m_{sus}$
is not strongly constrained but 27.5$^{\circ}$ is consistent with a number of $m_{sus}/m_e$ values (e.g., 3.99, 4.87, and 5.76), which are
all close to the measured thermodynamic mass $m^*$.

The overall suppression of the dHvA amplitude with $\theta$ is controlled by a combination of impurity scattering, doping inhomogeneity and
mosaic spread.  It is difficult to disentangle the three effects as they all have similar $B$ dependence over the limited field range where
the oscillations are observable. However, by assuming that each term alone is responsible for the damping we can set upper limits on their
individual magnitudes. The Dingle damping term due to impurity scattering is given by $R_D= \exp[-\sqrt{2\pi^2\hbar F_0/e}/ (\ell_0 B \cos
\theta)]$ while the damping term due to doping inhomogeneity or mosaic spread is $R_{p}$ = $\exp[-(\pi \alpha / B \cos \theta)^2]$ (where
$\alpha = \pi \hbar \delta p / e a^2$ or $F_0 \tan \theta \delta \theta$ respectively -- here we have assumed a Gaussian distribution for
both) \cite{Springford}. Typical fits for Tl10Ka and Tl10Kb, using $\delta p = 0.0025$, $\delta \theta$ = 0 and $\ell_0 = \infty$, are
shown as solid lines in Fig.~\ref{Fig3}. We find that $\delta p$, the spread in $p$, must be less than $0.0025(5)$ for all samples. This is
an important result as it demonstrates that the doping distribution in OD Tl2201 is negligibly small on the length scale of $\ell_0 >
400$\,\AA. Indeed, were $\delta p>$ 0.005, {\it all} dHvA phenomena would be strongly damped and therefore rendered unobservable within our
current experimental noise floor.

The overall $\theta$ dependence of the amplitude is difficult to model precisely as it depends on knowledge of the mosaic distribution in
the crystal as well as the $\theta,\varphi$ dependence of the scattering rate. In particular, we observed a strong asymmetry in the angular
dependence for $\varphi \sim 0^\circ$ that is not understood. This asymmetry is much weaker for $\varphi \sim 45^\circ$ and importantly,
the positions of the minima, which determine $k_{21}$, are symmetric with respect to $\pm \theta$.

Collectively, these findings rule out the notion that coexisting hole-rich and hole-poor regions (of order the coherence length) are the
origin of the decrease in $n_s$ in OD Tl2201 \cite{Uemura01, Uemura93}. In the alternative, pair-breaking scenario, the rapid loss of
superfluid density is attributed to a crossover from weak to strong pair breaking with overdoping \cite{Niedemayer93}. According to our
dHvA data, $\ell_0$ is relatively insensitive to carrier concentration and recent high-field transport measurements on LSCO indicate that
the residual in-plane resistivity $\rho_{ab}(0)$ is roughly constant across the entire OD region of the phase diagram \cite{Cooper09}. Thus
for the crossover from clean to dirty limit superconductivity to be realized, overdoping must be accompanied by a marked reduction in the
strength of the pairing interaction \cite{Niedemayer93, Storey07}, as implied by the observed correlation between $T_c$ and the magnitude
of the (anisotropic) $T$-linear scattering rate in OD LSCO \cite{Cooper09} and Tl2201 \cite{Abdel-Jawad07}.

The SC phase diagram of OD Tl2201 is compared with that of LSCO in the inset to Fig.~\ref{Fig2}. The dashed line is the \lq universal'
parabola \cite{Tallon95}, scaled to $T_c^{\rm max}$, while the solid line is the corresponding $T_c(p)$ curve for Tl2201, as determined by
this study (black squares). (Here we have assumed that $T_c^{\rm max}$ remains at $p = 0.16$.) Extrapolation of the solid line in
Fig.~\ref{Fig2} to $T_c = 0$ implies that superconductivity in Tl2201 will disappear at $p = 0.31$.  In LSCO, the $T$-linear term in
$\rho_{ab}$ persists to $x = 0.29$, i.e. outside the LSCO SC dome \cite{Cooper09}, implying that pairing may still be active there. Indeed,
comparison of the impurity scattering rate with $\Delta_0$ suggests that the parabolic tail-off of $T_c(p)$ in LSCO could be attributable
to the same pair-breaking effects that lead to the reduction in $n_s/m^*$. In LSCO ($x = 0.29$), $\rho_{ab}(0) \sim 18$\,$\mu\Omega$cm
\cite{Cooper09}. Taking FS parameters for OD LSCO from ARPES \cite{Yoshida07}, we obtain a transport (i.e.~large-angle) scattering rate
$\hbar/\tau_0 \sim$ 10\,meV that is much larger than the BCS weak coupling value $\Delta_0 = 2.14 k_B T_c \sim 2$\,meV, for  $T_c \simeq
10$\,K. For Tl2201 with $T_c = 10$\,K, $\rho_{ab}(0) \sim 6$\,$\mu\Omega$cm \cite{Proust02} and correspondingly, $\hbar/\tau_0 \sim 3$\,meV
$\simeq\Delta_0$.

In conclusion, detailed angle-dependent QO experiments indicate that there is no phase separation in OD Tl2201 over a length scale of
hundreds of Angstroms. All indicators suggest that the physical properties of OD Tl2201 are determined by a single, spatially homogeneous
Fermi liquid electronic state that is now very well-characterized. It would appear that static nanoscale inhomogeneity and phase separation
are not {\it generic} features of cuprates {\it in any region of the phase diagram}. We therefore conclude that pair breaking (possibly
enhanced by the effect of a pairing interaction that is highly anisotropic) is responsible for the loss of superfluid density in OD Tl2201
and probably for the disappearance of superconductivity in LSCO below $p_c$ = 0.31. The underlying reason for this appears to be the rapid
fall in the strength of the pairing interaction on the OD side \cite{Niedemayer93,Storey07}. This, and the absence of any significant
renormalization near the Fermi level support a purely magnetic or electronic mechanism. Our findings also provide further experimental
evidence that superconductivity persists to much higher doping levels than the normal state pseudogap. We stress here that the closure of
the pseudogap is not field-induced, since the FS parameters found here are entirely consistent with zero-field transport
\cite{Mackenzie96b}, thermodynamic \cite{Wade94} and spectroscopic \cite{Plate05} data.

We thank L.~Balicas, A.~I.~Coldea, C.~Proust, D.~Vignolles, B.~Vignolle, I.~Kokanovi\'{c}, A.~P.~Mackenzie, D.~A.~Bonn, W.~N.~Hardy,
R.~Liang and B.~J.~Ramshaw for their contributions to this project. This work was supported by the EPSRC (UK), the Royal Society and a
co-operative agreement between the State of Florida and NSF. 



\vskip -1cm
\begin{thebibliography}{32}

\bibitem{Takagi89} H. Takagi {\it et~al.}, Phys. Rev. B {\bf 40}, 2254 (1989).
\bibitem{Tallon95} J. L. Tallon {\it et~al.}, Phys. Rev. B {\bf 51}, 12911 (1995).
\bibitem{Uemura01} Y. J. Uemura, Solid State Comm. {\bf 120}, 347 (2001).
\bibitem{Cren00} T. Cren {\it et~al.}, Phys. Rev. Lett. {\bf 84}, 147 (2000).
\bibitem{McElroy05} K. McElroy {\it et~al.} Science {\bf 309}, 1048 (2005).
\bibitem{EmeryKivelson93} V. J. Emery and S. A. Kivelson, Physica (Amsterdam) {\bf 209C}, 597 (1993).
\bibitem{Bobroff02} J. Bobroff {\it et~al.}, Phys. Rev. Lett. {\bf 89}, 157002 (2002).
\bibitem{Loram04} J. W. Loram {\it et~al.}, Phys. Rev. B {\bf 69}, 060502(R) (2004).
 \bibitem{Niedemayer93} Ch. Niedermayer {\it et~al.}, Phys. Rev. Lett. {\bf 71}, 1764 (1993).
\bibitem{Uemura93} Y. J. Uemura  {\it et~al.}, Nature (London) {\bf 364}, 605 (1993).
\bibitem{Tanabe05} Y. Tanabe {\it et~al.}, J. Phys. Soc. Japan {\bf 74}, 2893 (2005).
\bibitem{Wang07} Y. Wang {\it et~al.}, Phys. Rev. B {\bf 76}, 064512 (2007).
\bibitem{Tyler97} A. W. Tyler, PhD. Thesis, Cambridge University (1997).
\bibitem{Vignolle08} B. Vignolle  {\it et~al.}, Nature (London) {\bf 455}, 952 (2008).
\bibitem{angleerrornote} The difference in $F_0$ between the two 10\,K samples could be due to a small $\sim 5^\circ$ misalignment of the sample along the axis perpendicular to the rotation axis.
\bibitem{Mackenzie96b} A. P. Mackenzie {\it et~al.}, Phys. Rev. B {\bf 53}, 5848 (1996).
\bibitem{Shoenberg} D. Shoenberg, {\em Magnetic Oscillations in Metals}, Cambridge University Press, Cambridge (1984).
\bibitem{Wasserman} A. Wasserman {\it et al.}, J. Phys. Cond. Matt {\bf 3}, 5335 (1991).
\bibitem{Mackenzie96a} A. P. Mackenzie {\it et~al.}, Physica (Amsterdam) {\bf 263C}, 510 (1996).
\bibitem{Wade94} J. M. Wade {\it et~al.}, J. Supercon. {\bf 7}, 261 (1994).
\bibitem{RourkeNJP} P. M. C. Rourke {\it et~al.}, (unpublished).
\bibitem{Sahrakorpi} S. Sahrakorpi {\it et al.},  Physica (Amsterdam) {\bf C460-462}, 428 (2007).
\bibitem{Plate05} M. Plat\'e  {\it et~al.}, Phys. Rev. Lett.  {\bf 95}, 077001 (2005).
\bibitem{Bergemann03} C. Bergemann {\it et~al.}, Adv. Phys. {\bf 52}, 639 (2003).
\bibitem{Hussey03} N. E. Hussey  {\it et~al.}, Nature (London) {\bf 425}, 814 (2003).
\bibitem{actualvalues} The parameters $k_{40}$, $k_{61}$ and $k_{101}$ are not well determined by our dHvA measurements and are fixed at the values determined by ADMR \cite{Abdel-Jawad07},
e.g. for $T_c$=10\,K we used $k_{40}/k_{00}=-0.0315$, $k_{61}/k_{21}=0.71$, $k_{101}/k_{21}=-0.25$.
\bibitem{Springford} {\it Electrons at the Fermi Surface}, edited by by M. Springford (Cambridge University Press 1980).
\bibitem{actualangles} Azimuthal angles of $\varphi=6^\circ$ and $45^\circ$ were assumed for the fits.
\bibitem{Cooper09} R. A. Cooper {\it et~al.}, Science {\bf 323}, 603 (2009).
\bibitem{Storey07} J. G. Storey {\it et~al.}, Phys. Rev. B {\bf 76}, 174522 (2007).
\bibitem{Abdel-Jawad07} M. Abdel-Jawad {\it et~al.}, Phys. Rev. Lett.  {\bf 99}, 107002 (2007).
\bibitem{Yoshida07} T. Yoshida {\it et~al.}, J. Phys. Cond. Matter {\bf 19}, 125209 (2007).
\bibitem{Proust02} C. Proust {\it et al.}, Phys. Rev. Lett. {\bf 89}, 147003 (2002).

\end{thebibliography}
\end{document}